\documentstyle[12pt]{article}

\begin{document}

\begin{titlepage}

\begin{center}
{\LARGE\bf{Statistical Operator for Electroweak Baryogenesis}}

\bigskip
\bigskip
\bigskip

{\large J. B. Bronzan\footnote{electronic mail: bronzan@physics.rutgers.edu.}}

\bigskip
Department of Physics and Astronomy\\
Rutgers University\\
Piscataway, New Jersey 08855-0849\\

\end{center}

\bigskip

\abstract{
\noindent
In electroweak baryogenesis, a domain wall between the spontaneously broken 
and unbroken phases acts as a separator of baryon (or lepton) number, 
generating a baryon asymmetry in the universe.  If the wall is thin relative 
to plasma mean free paths, one computes baryon current into the broken 
phase by determining the quantum mechanical transmission of plasma 
components in the potential of the spatially changing Higgs VEV.  
We show that baryon current can also be obtained using a statistical 
density operator.  This new formulation of the problem provides 
a consistent framework for studying the influence of 
quasiparticle lifetimes on baryon current.  We show that when the plasma 
has no self-interactions, familiar results are reproduced.  When plasma 
self-interactions are included, the baryon current into the broken phase 
is related to an imaginary time temperature Green's function.}

\vfill

\leftline{Rutgers-97-65}
\leftline{July, 1997}

\end{titlepage}

\section{Introduction}
\label{Introduction}
\setcounter{equation}{0}
     During the electroweak phase transition, baryogenesis can occur in the 
minimal standard model and its extensions (more Higgs doublets, supersymmetry, 
etc.), provided the phase transition is strongly first order\cite{Ru}.  
Bubbles of 
broken symmetry (true) vacuum sweep through the unbroken symmetry (unstable) 
vacuum.  The computation of the ratio of baryon density to entropy density 
(measured to be $\sim 10^{-10}$\cite{steig}) proceeds differently if the 
bubble wall is thin\cite{Co1,Co2,Ne} or thick\cite{McL,Di,Co3} relative to 
the mean free paths of constituents of the quark-lepton-boson plasma.  
In this paper we will be concerned with the thin wall case, where one 
computes the transmission of plasma quasiparticles past the bubble wall.  
Because CP and C are violated, the transmission of baryon 
number (or its lepton surrogate) differs from transmission of antibaryon 
number.  This asymmetry drives the baryon asymmetry of the universe, although 
a number of other physical effects are important in arriving at the final 
result.

     Farrar and Shaposhnikov used this framework to estimate the 
baryon asymmetry produced by the minimal standard model, with CP violation 
originating in the CKM matrix\cite{farrar1}.  They concluded the 
experimentally observed number could be generated in this way.  This result 
was criticized by Gavela {\it et.al.}, who obtained an asymmetry about 
$10^{-13}$ relative to that computed by Farrar and Shaposhnikov\cite
{gavela}.  These authors state that the main reason for the discrepancy 
is the very short lifetimes of quasiparticles in the hot plasma 
($\Gamma\sim 19$ GeV for quarks).  On the other hand, in the work of 
Gavela {\it et.al.}, quasiparticle damping is incorporated by giving 
quasiparticles a non-Hermitian effective Lagrangian.  This falls short 
of a complete many-body treatment of the self-interacting plasma moving 
past the Higgs VEV (vacuum expectation value) domain wall.

     In this paper we develop an alternate framework for computing 
currents which may make it possible to reduce the number of {\it ad hoc} 
assumptions in calculations.  We advocate using a statistical operator that 
specifies the thermal system consisting of the hot, self-interacting plasma 
with the VEV domain wall moving through it.  In principle, the new 
framework does not require the introduction of quasiparticles at all.  It 
could be used even when lifetimes are so short that quasiparticles cease 
to be useful degrees of freedom.  In practice, one must be able to accurately 
take account of the plasma self-interactions, and this usually 
means working in terms of physical excitations, using perturbation theory 
to account for interactions.  Finite quasiparticle lifetimes and mean free 
paths are present in the statistical mechanics approach as a consequence of 
self-interactions in the plasma, and when quasiparticles decay, baryon 
current borne by decay products is included.

     Two Lorentz frames are involved in the problem.  The ``laboratory'' frame 
is the rest frame of the plasma, denoted by $P$.  The domain wall moves 
through frame $P$ at speed $v$.  (We take the domain wall to be planar, 
thereby ignoring the curvature of the bubble wall.)   The second frame is 
the rest frame of the domain wall, denoted $V$.  In frame $V$, the plasma 
streams by the stationary VEV wall, from the unbroken into the broken phase.  
We will take the unbroken phase to lie at $z<0$, and the broken phase at 
$z>0$.  We assume, with references~\cite{farrar1,gavela}, that the latent heat 
associated with the phase transition can be ignored.  Then there is no 
energy source at the bubble wall, and we are dealing with a closed system 
described by a Hamiltonian.  Like these authors, we also discount turbulence 
so that the system is stationary in frame $V$.  Taken together, these 
observations imply that the plasma is described by observers in $V$ by a 
time-independent statistical operator $\rho$.  Because the thermal bath in 
contact with the gas is at rest in frame $P$, the density matrix is 
\begin{equation}
\rho = \exp [-\beta K_P],
\end{equation}
where $K_P=H_P-\mu N$, and $H_P$ is the Hamiltonian in $P$.  What forces us 
to deal with two frames is that while $K_P$ determines $\rho$, $K_P$ 
is time-independent only in frame $V$.  [Recall that to recast the 
computation of the baryon current as a conventional problem in 
equilibrium statistical mechanics, $\rho$ must be time-independent.]

     In frame $V$ there is a conserved baryon current $\hat J_\mu ({\bf x},t)$ 
whose form and time-dependence are determined by $H_V$, the Hamiltonian of 
the system in $V$.\footnote{$\hat{\cal O}$ is used when we wish to 
distinguish an operator from a related c-number $\cal O$.}  
The baryon current past the VEV wall, measured in $V$, is 
\begin{equation}
\label{eq:thermcurr}
J(z,t)={{Tr[\rho \hat J_3({\bf x},t)]}\over{Tr[\rho]}}.
\end{equation}
This expression appears simpler than it is because one Hamiltonian appears 
in $\rho$, a second Hamiltonian controls time evolution, and 
the two do not commute.  As a result, $J$ generally depends on $z$ and $t$.  
However, as $t\rightarrow\infty$, transients decay and the current approaches 
a limit that is 
constant in time and space.  In the case of a plasma without 
self-interactions, the late time current is what is expected on the basis 
of simple physical arguments.  The resulting formula (Eq.~\ref
{eq:freecurrent}) has been used in previous work, either implicitly or 
explicitly.
 
     The matter content of the plasma in the epoch of baryogenesis depends on 
whether the standard model or one of its extensions is under study.  Even 
were we to choose a specific case, the inclusion of many species would tend 
to obscure the gist of our work.  In a similar vein, implementation of 
realistic baryon-nonconserving processes would introduce complexity and 
detail that are undesirable.  What is illuminating is to study a simple 
system for which the statistical matrix approach is developed in a way that 
can be generalized to other plasmas.  In this paper we assume a single spin 
zero baryon-antibaryon species is present in the plasma.  The Higgs VEV 
introduces a $z$-dependent mass $m(z)$ for these bosons.  We assume that the 
VEV jumps to its broken symmetry value only in the region $0<z<L$.  That is, 
in this interval, $m(z)\sim M$, while elsewhere $m(z)\sim m$, and $m<M$.  
Having $m(+\infty )=m(-\infty )$ has the advantage of simplifying the 
asymptotic form of basis functions that appear extensively in the 
calculations, and it is for this reason that the width of the broken 
symmetry phase is finite.  Since $L$ can be as large as we wish, its 
finitude has no physical significance because tunneling through the broken 
symmetry phase can be made as small as we wish by increasing $L$.  

     If nothing further were done, we would find $J=0$ because while there 
would be a baryon current past the domain wall, there would 
be an equal and canceling antibaryon current.  We produce a net baryon 
current by introducing a chemical potential that enhances the baryon 
density relative to the antibaryon density.  It plays the role of baryon 
non-conserving processes in our model.  In order for the Bose distribution 
functions to be finite, the chemical potential must lie in the range 
$-m<\mu<m$.  This restriction is what requires us to choose $m\ne 0$ in the 
``unbroken symmetry'' phase in our model.

     In Section 2 we consider the diagonalization of $K_P$, $H_P$, $K_V$ and 
$H_V$, in the case where the plasma has no self-interactions.  This allows 
us to compute the baryon current of such a plasma in Section 3.  
In the limit $t\rightarrow\infty$, the 
baryon current we find is the expected one.  In Section 4 we consider the 
case of a plasma with self-interactions.  The 
objective there is to express the current in terms of an {\it imaginary 
time} temperature Green's function in equilibrium many-body theory.  It is 
these Green's functions that have simple expansions in perturbation 
theory, and to which the formalism and methods of many-body theory apply.  
In Section 4 a number of complications come together: 
the difference between $H_P$ and $H_V$, the need to take $t\rightarrow\infty$, 
and the requirement of relating real time and imaginary time temperature 
Green's functions.  We succeed in reconciling these strands only by 
working to first order in the plasma speed $v$.  (In Section 3 our results 
for a plasma without self-interactions hold for $0<v<1$.)  We argue that 
the baryon current $J(v)$ should be odd in $v$, so this restriction is less 
stringent than it might seem.  Extension of our results to $O(v^3)$ is 
conceivable, but would be complicated.  At the end of Section 4 we have a 
specific imaginary time temperature Green's function which gives the 
baryon current to $O(v)$.  The effect of quasi-particle decay on baryon 
current is now accounted for by assessing the effect of plasma 
self-interactions on this Green's function.  In particular, contributions 
to the boson propagator where a boson decays on one side of the VEV wall and 
reassembles on the other correspond to contributions to the baryon current 
from quasi-particle decay products.

\section{Free Fields}
\setcounter{equation}{0}

     To show that a statistical formulation of the calculation of baryon 
current works, we first consider a plasma that has no self-interactions, but 
full interaction with the VEV wall.  As a preparation, we formulate and 
diagonalize the four operators $K_{P0}$, $H_{P0}$, $K_{V0}$, and 
$H_{V0}$.  $H_{V0}$, the free Hamiltonian in frame $V$, differs from the 
conventional free boson Hamiltonian only through its spatially varying mass 
term:
\begin{equation}
\label{eq:HV0}
H_{V0}=\int d^3x\left [\hat\pi\hat\pi^\dagger+\nabla\hat\phi^\dagger
{\bf\cdot}\nabla\hat\phi +m^2(z)\hat\phi^\dagger\hat\phi\right ].
\end{equation}
$H_{V0}$ generates $t$-displacements; we also 
need the momentum operator $P_{V,3}$, which generates  $z$-displacements:
\begin{equation}
\label{eq:PV3}
P_{V,3}=\int d^3x\left [{{\partial\hat\phi^\dagger}\over{\partial z}}
\hat\pi^\dagger +\hat\pi{{\partial\hat\phi}\over{\partial z}}\right ].
\end{equation}
Then for any Heisenberg operator $\cal O$, 
\begin{equation}
{{\partial{\cal O}}\over{\partial t}}=-i\left [{\cal O},H_{V0}\right ];
\qquad{{\partial{\cal O}}\over{\partial z}}=-i\left [{\cal O},P_{V,z}\right ].
\end{equation}
Observers in frame $P$ attribute primed coordinates to events, where\footnote
{$v$ is the relative speed between $P$ and $V$.  We use $\beta$ for $1/kT$.}
\begin{equation}
t'=\gamma (t-vz);\qquad z'=\gamma (z-vt).
\end{equation}
Thus, 
\begin{equation}
{{\partial{\cal O}}\over{\partial t'}}=\gamma{{\partial{\cal O}}\over
{\partial t}}+\gamma v{{\partial{\cal O}}\over{\partial z}}=
-i\left [{\cal O},\gamma H_{V0}+\gamma vP_{V,z}\right ].
\end{equation}
Since the generator of $t'$-displacements is $H_{P0}$, we have 
\begin{equation}
H_{P0}=\gamma H_{V0}+\gamma v P_{V,3}.
\end{equation}
(A similar formula holds for a plasma with self-interactions; it will 
be used in Section 4.)  Finally, the grand canonical Hamiltonians 
are $K_{V0}=H_{V0}-\mu N$ and 
$K_{P0}=H_{P0}-\mu N,$ where the baryon minus antibaryon number 
operator is 
\begin{equation}
\label{eq:N}
N=i\int d^3x\left [\hat\phi^\dagger\hat\pi^\dagger-\hat\pi\hat\phi\right ].
\end{equation}

     $K_{P0}$ is the key to diagonalizing these operators, since once 
$K_{P0}$ is in diagonal form, the remaining operators follow by setting 
$v=0$, or $\mu =0$, or both.  The construction begins by considering the 
Heisenberg equations of motion that are generated by $K_{P0}$: 
\begin{equation}
{{\partial^2\hat\phi}\over{\partial t^2}}=\gamma^2\nabla^2_\perp\hat\phi+
{{\partial^2\hat\phi}\over{\partial z^2}}+2\gamma v{{\partial^2\hat\phi}\over
{\partial t\partial z}}+2i\mu{{\partial\hat\phi}\over{\partial t}}
-2i\gamma v\mu {{\partial\hat\phi}\over{\partial z}}+\mu^2\hat
\phi-\gamma^2m^2(z)\hat\phi,
\label{eq:motion}
\end{equation}
\begin{equation}
\label{eq:momentum}
\hat\pi^\dagger ={1\over\gamma}{{\partial\hat\phi}\over{\partial t}}
-v{{\partial\hat\phi}\over{\partial z}}-i{\mu\over\gamma}\hat\phi.
\end{equation}
This equation implies that the current $\hat I_\mu$ is conserved, where
\begin{eqnarray}
\label{eq:current1}
\hat I_0 & = & i\hat\phi^\dagger\left ({{\stackrel{\leftrightarrow}
{\partial}}\over{\partial t}}-\gamma v{{\stackrel{\leftrightarrow}
{\partial}}\over{\partial z}}-2i\mu\right )\hat\phi,\nonumber \\
{\bf\hat I_\perp} & = & -i\gamma^2\hat\phi^\dagger\stackrel
{\longleftrightarrow}{\nabla_\perp}\hat\phi,\\
\hat I_3 & = & -i\hat\phi^\dagger\left ({{\stackrel{\leftrightarrow}
{\partial}}\over{\partial z}}+\gamma v{{\stackrel{\leftrightarrow}
{\partial}}\over{\partial t}}-2i\gamma v\mu\right )\hat\phi,\nonumber
\end{eqnarray}
and $\hat\phi$ satisfies Eq.~\ref{eq:motion}. (When we replace 
the free Hamiltonian $H_{P0}$ by the full Hamiltonian $H_P$, 
Eq.~\ref{eq:motion} acquires additional interaction 
terms and counterterms on the right.  $\hat I_\mu$ is still conserved 
in the theory with self-interactions.)  Setting $v=\mu=0$, the conserved 
baryon current in frame $V$ and Eq.~\ref{eq:thermcurr} is
\begin{equation}
\label{eq:current2}
\hat J_0=i\hat\phi^\dagger {{\stackrel{\leftrightarrow}{\partial}}
\over{\partial t}}\hat\phi,\qquad {\bf\hat J}=-i\hat\phi^\dagger
\stackrel{\longleftrightarrow}{\nabla}\hat\phi.
\end{equation}
     
     We obtain basis states for the diagonalization of $K_{P0}$ by 
finding classical solutions of Eq.~\ref{eq:motion}.  If we write 
$\phi = \tilde\phi\exp [i\mu t]$, the chemical potential drops out, 
and $\tilde\phi$ satisfies
\begin{equation}
{{\partial^2\tilde\phi}\over{\partial t^2}}=\gamma^2\nabla^2_\perp\tilde\phi+
{{\partial^2\tilde\phi}\over{\partial z^2}}+2\gamma v{{\partial^2\tilde\phi}
\over{\partial t\partial z}}-\gamma^2m^2(z)\tilde\phi,
\end{equation}
By separation of 
variables, we have solutions $\tilde\phi=\exp (-iEt)P({\bf k,x})$ 
where
\begin{equation}
P({\bf k,x})={{\exp [i{\bf k_\perp\cdot x_\perp}+i\gamma vEz]}\over
{[2E(2\pi )^3]^{1/2}}}{\cal Z}_P(k_3,z),
\end{equation}
$E=\sqrt{{\bf k^2}+m^2}$, and
\begin{equation}
\label{eq:zeq}
{{d^2{\cal Z}_P(k_3,z)}\over{dz^2}}+\gamma^2[k_3^2+m^2-m^2(z)]{\cal Z}_P
(k_3,z)=0.
\end{equation}

     The current defined in Eq.~\ref{eq:current1} continues to be conserved 
when $\hat\phi$ is replaced by one classical solution, and $\hat\phi^\dagger$ 
by the complex conjugate of a second solution.  Consider the integral of the 
resulting charge density.  Introducing useful notation, we write the total 
charge as
\begin{equation}
(\phi ({\bf k}),\phi ({\bf k'}))=e^{i(E-E')t}
(P({\bf k}),P({\bf k'})),
\end{equation}
\begin{displaymath}
(P({\bf k}),P({\bf k'}))=\int d^3xP^*({\bf k},{\bf x})\left [
E+E'-i\gamma v{{\stackrel{\leftrightarrow}{\partial}}
\over{\partial z}}\right ]P({\bf k'},{\bf x})
\end{displaymath}
\begin{displaymath}
={{\gamma\delta ({\bf k_\perp}-{\bf k'_\perp})}\over{4\pi\sqrt{EE'}}}
\int dze^{i\gamma v(E'-E)z}
\left\{{\cal Z}_P^*(k_3,z)\left [\gamma (E+E')-iv{{\stackrel
{\leftrightarrow}d}\over{dz}}\right ]{\cal Z}_P(k'_3,z)\right\}.
\end{displaymath}
However, the total charge should be time independent, implying that
\hfill
\break 
$(P({\bf k}),P({\bf k'}))$ vanishes for $E\ne E'$. This makes 
$(P({\bf k}),P({\bf k'}))$ an appropriate inner product for the 
time-independent solutions.  

     Orthogonality can be verified directly.  It follows from 
Eq.~\ref{eq:zeq} that
\begin{equation}
{d\over{dz}}\left \{e^{i\gamma v(E'-E)z}{\cal Z}_P^*(k_3,z)
\left [{{\stackrel{\leftrightarrow}d}\over{dz}}\right ]{\cal Z}_P(k'_3,z)
\right \}
\end{equation}
\begin{displaymath}
=\gamma (E-E')e^{i\gamma v(E'-E)z}
\left\{{\cal Z}_P^*(k_3,z)\left [\gamma (E+E')-iv{{\stackrel
{\leftrightarrow}d}\over{dz}}\right ]{\cal Z}_P(k'_3,z)\right \}.
\end{displaymath}
Integrating this, the left side vanishes, so the inner product is zero 
unless $E=E'$ or $k_3=\pm k_3'$.

     When $k_3=\pm k_3'$, there is a delta function contribution to the 
inner product generated by the integral over $z$.  Its strength is determined 
by the asymptotic behavior of ${\cal Z}_P$, which we now specify.  At large 
$|z|$, solutions of Eq.~\ref{eq:zeq} are plane waves, and we adopt scattering 
boundary conditions.  For 
$k_3>0$,
\begin{eqnarray}
\label{eq:bcplus}
{\cal Z_P}(k_3,z)=\cases{\exp (i\gamma k_3z)+r_P(k_3)\exp (-i\gamma k_3
z),&$(z<<0)$,\cr 
t_P(k_3)\exp (i\gamma k_3z),&$(z>>L)$.\cr},
\end{eqnarray}
and for $k_3<0$,
\begin{eqnarray}
\label{eq:bcminus}
{\cal Z}_P(k_3,z)=\cases{t_P(k_3)\exp (i\gamma k_3z),&$(z<<0)$,\cr 
\exp (i\gamma k_3z)+r_P(k_3)\exp (-i\gamma k_3z),&$(z>>L)$.\cr}
\end{eqnarray}  
These scattering amplitudes satisfy ``unitarity'' relations that we use 
repeatedly.  
Eq.~\ref{eq:zeq} implies that the expressions
\begin{displaymath}
{\cal Z}_P^*(k_3,z)\left [{{\stackrel{\leftrightarrow}d}\over{dz}}\right ]
{\cal Z}_P(k'_3,z),\qquad{\cal Z}_P(k_3,z)\left [{{\stackrel
{\leftrightarrow}d}\over{dz}}\right ]{\cal Z}_P(-k_3,z)
\end{displaymath}
are independent of $z$ when $k'_3=\pm k_3$.  Evaluating the expressions at 
large positive and negative $z$, we obtain the unitarity relations
\begin{eqnarray}
\label{eq:relations}
|r_P(k_3)|^2+|t_P(k_3)|^2=1,\\
r^*_P(k_3)t_P(-k_3)+t^*_P(k_3)r_P(-k_3)=0,\nonumber\\
t_P(-k_3)=t_P(k_3).\nonumber
\end{eqnarray}
Using Eq.~\ref{eq:relations}, the complete orthogonality relation is
\begin{equation}
\label{eq:orthog}
(P({\bf k}),P({\bf k'}))=\gamma\delta ({\bf k}-{\bf k'}).
\end{equation}

     Solutions studied so far have $E>0.$  Related negative 
frequency solutions are the complex conjugates  
$\tilde\phi=\exp (+iEt)P^*({\bf k,x})$.  
The inner product of a negative energy solution with a positive energy 
solution is
\begin{equation}
((P^*({\bf k}),P({\bf k'}))=\int d^3xP({\bf k})\left [
E'-E-i\gamma v{{\stackrel{\leftrightarrow}{\partial}}
\over{\partial z}}\right ]P({\bf k'})
\end{equation}
\begin{displaymath}
={{\gamma\delta ({\bf k_\perp}+{\bf k'_\perp})}\over
{4\pi\sqrt{EE'}}}\int dze^{i\gamma v(E'+E)z}
\left\{{\cal Z}_P(k_3,z)\left [\gamma (E'-E)-iv{{\stackrel
{\leftrightarrow}d}\over{dz}}\right ]{\cal Z}_P(k'_3,z)\right\}.
\end{displaymath}
It can be shown that the integral over $z$ always vanishes.  The complete 
set of orthonormality relations is
\begin{equation}
(P({\bf k}),P({\bf k'}))=-(P^*({\bf k'}),P^*({\bf k}))=\gamma\delta 
({\bf k}-{\bf k'}),\qquad (P^*({\bf k}),P({\bf k'}))=0.
\end{equation}

     The quantum fields can be expanded in terms of these solutions.
\begin{equation}
\label{eq:phiheis}
\hat\phi ({\bf x},t)=e^{i\mu t}\int d^3k \left [e^{-iEt}
a({\bf k})P({\bf k},{\bf x})+e^{iEt}b^\dagger ({\bf k})
P^*({\bf k},{\bf x})\right ].
\end{equation}
Using Eq.~\ref{eq:momentum}, we find
\begin{equation}
\label{eq:piheis}
\hat\pi^\dagger ({\bf x},t)=-ie^{i\mu t}\int d^3k \left [e^{-iEt}
a({\bf k})\left ({{E}\over{\gamma}}-iv{\partial\over{\partial z}}\right )
P({\bf k},{\bf x})\right. 
\end{equation}
\begin{displaymath}
\left. -e^{iEt}b^\dagger ({\bf k})
\left ({{E}\over{\gamma}}+iv{\partial\over{\partial z}}\right )
P^*({\bf k},{\bf x})\right ].
\end{displaymath}
Now set $t=0$ to obtain Schr\"odinger picture fields.  Using orthonormality, 
we can project out the operator coefficients
\begin{equation}
a({\bf k})=\int d^3x\left [\hat\phi ({\bf x})\left ({E\over\gamma}
+iv{{\partial}\over{\partial z}}\right )P^*({\bf k},{\bf x})+i\hat\pi^\dagger 
({\bf x})P^*({\bf k},{\bf x})\right ],
\end{equation}
\begin{displaymath}
b^\dagger ({\bf k})=\int d^3x\left [\hat\phi ({\bf x})\left ({E\over
\gamma}-iv{{\partial}\over{\partial z}}\right )P({\bf k},{\bf x})-i\hat
\pi^\dagger ({\bf x})P({\bf k},{\bf x})\right ].
\end{displaymath}
Canonical commutation relations for the fields imply that these operators 
all commute, except
\begin{equation}
\left [a({\bf k}),a^\dagger ({\bf k'})\right ]=\left [b({\bf k}),b^\dagger 
({\bf k'})\right ]=\delta ({\bf k}-{\bf k'}).
\end{equation}
These operators destroy baryons ($a$) and antibaryons ($b$).

     The reason our construction leads to a basis that diagonalizes 
$K_{P0}$ is that, by Eqs.~\ref{eq:phiheis} and \ref{eq:piheis}, the time 
dependence of the destruction and creation operators is sinusoidal.  This 
implies 
\begin{equation}
\left [a({\bf k}),K_{P0}\right ]=(E-\mu)a({\bf k}),\qquad
\left [b({\bf k}),K_{P0}\right ]=(E+\mu)b({\bf k}).
\end{equation}
For these equations to hold we must have, aside from a normal ordering 
constant that is irrelevant in this problem,
\begin{equation}
K_{P0}=\int d^3k\left [\left (E-\mu\right )a^\dagger ({\bf k})a({\bf k})
+\left (E+\mu\right )b^\dagger ({\bf k})b({\bf k})\right ].
\end{equation}
This equation may be derived directly by substituting the expansions for the 
Schr\"odinger picture operators into Eqs.~\ref{eq:HV0}, \ref{eq:PV3} and 
\ref{eq:N}.

     When $v=0$, $K_{P0}$ becomes $K_{V0}$, and the basis functions 
$P({\bf k},{\bf x})$ are replaced by
\begin{equation}
V({\bf k},{\bf x})={{\exp [i\bf k_\perp\cdot x_\perp]}\over{[2E(2\pi)^3
]^{1/2}}}{\cal Z}_V(k_3,z),
\end{equation}
\begin{displaymath}
{{d^2{\cal Z}_V(k_3,z)}\over{dz^2}}+[k_3^2+m^2-m^2(z)]{\cal Z}_V(k_3,z)=0.
\end{displaymath}
The fields may also be expanded in this basis:
\begin{equation}
\label{eq:fieldinV}
\hat\phi ({\bf x},t)=e^{i\mu t}\int d^3k \left [e^{-iEt}
A({\bf k})V({\bf k},{\bf x})+e^{iEt}B^\dagger ({\bf k})
V^*({\bf k},{\bf x})\right ],
\end{equation}
\begin{displaymath}
\hat\pi^\dagger ({\bf x},t)=-ie^{i\mu t}\int d^3k E\left [e^{-iEt}
A({\bf k})V({\bf k},{\bf x})-e^{iEt}B^\dagger ({\bf k})
V^*({\bf k},{\bf x})\right ].
\end{displaymath}
The new operators $A$ and $B$ also have the commutation relations to destroy 
baryons and antibaryons.  The operators $K_{V0}$ and $H_{V0}$ are diagonal 
when expressed in terms of these operators:
\begin{eqnarray}
K_{V0}=&\int& d^3k\left [\left (E-\mu\right )A^\dagger ({\bf k})
A({\bf k})+\left (E+\mu\right )B^\dagger ({\bf k})B({\bf k})\right ],
\nonumber\\
H_{V0}=&\int& d^3kE\left [A^\dagger ({\bf k})A({\bf k})+
B^\dagger ({\bf k})B({\bf k})\right ].
\end{eqnarray}
\vfill
\eject

\section{Current of Free Baryons}
\setcounter{equation}{0}

    For free baryons, the obstacle to the evaluation of 
Eq.~\ref{eq:thermcurr} is that $\rho$ 
is expressed in terms of frame $P$ operators $a,\dots$, while the Heisenberg 
picture operator $\hat J_3({\bf x},t)$ is expressed in terms of frame 
$V$ operators $A,\dots$:
\begin{eqnarray}
\hat J_3({\bf x},t)&=-i\int d^3kd^3k'\left [e^{i(E-E')t}
A^\dagger ({\bf k})A ({\bf k'})V^*({\bf k},{\bf x})\left ({{\stackrel
{\leftrightarrow}\partial}\over{\partial z}}\right )V({\bf k'},{\bf x})\right.
\nonumber\\
&+e^{i(E+E')t}A^\dagger ({\bf k})B^\dagger ({\bf k'}) V^*({\bf k},
{\bf x})\left ({{\stackrel{\leftrightarrow}\partial}\over{\partial z}}
\right )V^*({\bf k'},{\bf x})\\
&+e^{i(-E-E')t} B ({\bf k})A ({\bf k'})
V ({\bf k},{\bf x})\left ({{\stackrel{\leftrightarrow}\partial}\over
{\partial z}}\right )V({\bf k'},{\bf x})\nonumber\\
&+e^{i(-E+E')t} B ({\bf k})B^\dagger ({\bf k'})
V ({\bf k},{\bf x})\left.\left ({{\stackrel{\leftrightarrow}\partial}\over
{\partial z}}\right ) V^*({\bf k'},{\bf x})\right ].\nonumber
\end{eqnarray}
Because $E>0,$ $\forall k_3$, oscillations must suppress the 
contributions of the two middle terms as $t\rightarrow\infty$.  To evaluate 
the contributions of the remaining terms, we must write the 
$A,\dots$ in terms of the $a,\dots$.  Begin with the projections
\begin{eqnarray}
A({\bf k})&=&\int d^3x\left [\hat\phi ({\bf x})EV^*({\bf k},{\bf x})
+i\hat\pi^\dagger ({\bf x})V^*({\bf k},{\bf x})\right ],\\
B^\dagger ({\bf k})&=&\int d^3x\left [\hat\phi ({\bf x})EV({\bf k},
{\bf x})-i\hat\pi^\dagger ({\bf x})V({\bf k},{\bf x})\right ].\nonumber
\end{eqnarray}
Substitute expansions \ref{eq:phiheis} and \ref{eq:piheis}, 
evaluated at $t=0$:
\begin{eqnarray}
A({\bf k})=\int_{-\infty}^{\infty} dk_3'\left [
a({\bf k_\perp}+\hat zk_3')R_1(k_\perp,k_3,k_3')\right.\nonumber\\
\left. +b^\dagger ({\bf -k_\perp}+\hat zk_3')R_2(k_\perp,k_3,k_3')
\right ],\\
B^\dagger({\bf k})=\int_{-\infty}^{\infty} dk_3'\left [
a({-\bf k_\perp}+\hat zk_3')R_2^*(k_\perp,k_3,k_3')\right.\nonumber\\
\left. +b^\dagger ({\bf k_\perp}+\hat zk_3')R_1^*(k_\perp,k_3,k_3')
\right ],\nonumber
\end{eqnarray}
where
\begin{equation}
\label{eq:R}
R_1(k_\perp,k3,k3')=\int_{-\infty}^{\infty}{dze^{i\gamma vE'z}
\over{4\pi\sqrt{EE'}}}{\cal Z}_V^*(k_3,z)
\left (E+\gamma E'-iv{d\over{dz}}\right ){\cal Z}_P(k_3',z),
\end{equation}
\begin{displaymath}
R_2(k_\perp,k3,k3')=\int_{-\infty}^{\infty}{dze^{-i\gamma vE'z}
\over{4\pi\sqrt{EE'}}}{\cal Z}_V^*(k_3,z)
\left (E-\gamma E'+iv{d\over{dz}}\right ){\cal Z}_P^*(k_3',z).
\end{displaymath}
One of the traces we need for the baryon current is
\begin{eqnarray}
\label{eq:trace}
&Tr\left [\rho A^\dagger ({\bf k})A({\bf k'})\right ]=\int_{-\infty}^\infty 
dq_3dq_3'\\
&\times\left \{R_1^*(k_\perp,k_3,q_3)R_1(k_\perp',k_3',q_3')Tr\left [\rho 
a^\dagger ({\bf k_\perp}+\hat zq_3)a({\bf k_\perp'}+\hat zq_3')\right ]
\right.\nonumber\\
&\left.+R_2^*(k_\perp,k_3,q_3)R_2(k_\perp',k_3',q_3')Tr\left [\rho 
b({\bf -k_\perp}+\hat zq_3)b^\dagger ({-\bf k_\perp'}+\hat zq_3')\right ]
\right \}.\nonumber
\end{eqnarray}
The traces in Eq.~\ref{eq:trace} are standard~\cite{wick}, leading to the 
expression
\begin{eqnarray}
\label{eq:S}
{{Tr\left [\rho A^\dagger ({\bf k})A({\bf k'})\right ]}\over
{Tr\left [\rho\right ]}}=\delta ({\bf k_\perp}-{\bf k_\perp'})
S_A(k_\perp,k_3,k_3'),\\
S_A(k_\perp,k_3,k_3')=\int_{-\infty}^\infty dq_3\left \{{{R_1^*(k_\perp,k_3,
q_3)R_1(k_\perp,k_3',q_3)}\over{e^{\beta 
(E-\mu)}-1}}\right.\nonumber\\
+\left. R_2^*(k_\perp,k_3,q_3)R_2(k_\perp,k_3',q_3)\left [{1\over{e^{\beta 
(E+\mu)}-1}}+1\right ]\right \}.\nonumber
\end{eqnarray}
In this equation, $E=\sqrt{k_\perp^2+q_3^2+m^2}$.  There is a similar 
formula for 
\hfill
\break
$Tr[\rho B({\bf k})B^\dagger ({\bf k'})]$.  $S_B$ differs from 
$S_A$ by the interchange $R_1\leftrightarrow R_2.$

     The baryon current at late time is
\begin{eqnarray}
J(z,t)=\int {{d^2k_\perp dk_3dk_3'}\over{2(2\pi)^3\sqrt{EE'}}}
\left [{\cal Z}_V^*(k_3,z)\left (-i{{\stackrel{\leftrightarrow}d}\over{dz}}
\right ){\cal Z}_V(k_3',z)\right ]\\
\times\left [e^{i(E-E')t}S_A(k_\perp,k_3,k_3')-e^{-i(E-E')t}S_B(k_\perp,k_3,
k_3')\right ].\nonumber
\end{eqnarray}
To extract the limiting current, consider the contribution of the term 
proportional to $S_A$.  First integrate over $k_3$, for fixed values of 
${\bf k_\perp}$ and $k_3'$.  The $k_3$ integration is initially 
along the real axis of the complex $k_3$ plane.  Deform the contour to $C$, 
which runs from $k_3=-\infty$ slightly below the real axis (in the third 
quadrant), crosses the real axis at $k_3=0$, and continues to $k_3=+\infty$ 
slightly above the real axis (in the first quadrant).  Except at the origin, 
contributions on $C$ are exponentially damped as $t\rightarrow\infty$, and at 
the origin contributions are suppressed by oscillation.  The contribution 
to the limiting value 
of $J(z,t)$ is therefore zero {\it unless} singularities are encountered 
when the contour is changed from the real axis to $C$.  A time-independent 
contribution arises if $S_A$ has a pole bordering the real axis, at 
$k_3=|k_3'|+i\eta$, or at $k_3=-|k_3'|-i\eta$, since at $k_3=\pm k_3'$, 
we have $E=E'$.  In fact $S_A$ as defined by Eq.~\ref{eq:S} has 
such poles.  Note that when $k_3=\pm k_3$, Eq.~\ref{eq:zeq} implies 
that $J$ is also independent of $z$.  

  By Eq.~\ref{eq:S}, poles in $S_A$ arise where poles of $R_i$ 
pinch the 
$q_3$ integration contour.  The $R_i$ have poles adjacent to 
the real $q_3$ axis owing to divergence of the integrals at $z=\pm\infty$ 
in Eq.~\ref{eq:R}.  
The positions and residues of these poles are determined by the asymptotic 
forms of ${\cal Z}_V$ and ${\cal Z}_P$, and therefore depend on the profile 
of the VEV wall only through the scattering amplitudes.

     The asymptotic behaviors change with the sign of the real parts of $k_3$, 
$k_3'$, and $q_3$, and many contributions must be added to determine the 
residues of the poles of $S_A$ at $k_3=\pm k_3'$.  Here we consider only 
two of the contributions to illustrate what is involved.

     When $k_3>0$ and $q_3<0$, the poles of $R_1$ arising at $z=-\infty$ are 
the same as those of
\begin{eqnarray}
R_1(k_\perp,k_3,q_3)&\sim&\int_{-\infty}^0{{dze^{\eta z+i\gamma vE_qz}}
\over{4\pi\sqrt{E_kE_q}}}\left [e^{ik_3z}+r_V(k_3)
e^{-ik_3z}\right ]^*\nonumber\\
&&\times\left (E_k+\gamma E_q
-iv{d\over{dz}}\right )t_P(q_3)e^{i\gamma q_3z}\\
&\sim&{{t_P(q_3)(E_k+\gamma E_q+\gamma vq_3)}\over{4i\pi
\sqrt{E_kE_q}}}\left [{1\over{\gamma q_3+\gamma vE_q-k_3
-i\eta}}\right.\nonumber\\
&&+\left.{{r_V^*(k_3 )}\over{\gamma q_3+\gamma vE_q+k_3-i\eta}}
\right ].\nonumber
\end{eqnarray}
The first denominator vanishes at $q_3=\gamma k_3-\gamma vE_k$, and the 
second at $q_3=-\gamma k_3-\gamma vE_k$.  We rewrite the terms to 
exhibit poles in the $q_3$ plane.
\begin{eqnarray}
R_1(k_\perp,k_3,q_3)&\sim&{{t_P(\gamma k_3-\gamma vE_k)}\over{2i\pi 
(q_3-\gamma k_3+\gamma vE_k-i\eta )}}\sqrt{{{\gamma E_k-\gamma 
vk_3}\over E_k}}\nonumber\\
&&+{{t_P(-\gamma k_3-\gamma vE_k)r^*_V(k_3)}\over{2i\pi (q_3+\gamma k_3
\gamma+vE_k-i\eta )}}\sqrt{{{\gamma E_k+\gamma vk_3}\over E_k}}.
\end{eqnarray}
Substituting either of these contributions into Eq.~\ref{eq:S} produces a 
pole at $k_3=k_3'$.  The two contributions are
\vfill
\eject
\begin{eqnarray}
\lefteqn{S_A(k_\perp,k_3,k_3')\sim}\\
\nonumber\\
&&{{|t_P(\gamma k_3'-\gamma vE_k')|^2(\gamma
E_k'-\gamma vk_3')\theta (\gamma v\sqrt{k_\perp^2+m^2}-k_3')}\over
{2\pi iE_k'(\gamma k_3-\gamma vE_k-\gamma k_3'+\gamma v
E_k'-i\eta )[\exp \beta(\gamma E_k'-\gamma vk_3'-\mu)-1]}}
\nonumber\\
\nonumber\\
\nonumber\\
&&+{{|t_P(\gamma k_3'+\gamma vE_k')|^2|r_V(k_3')|^2(\gamma E_k'
+\gamma vk_3')}\over
{2\pi iE_k'(-\gamma k_3-\gamma vE_k+\gamma k_3'+\gamma v
E'-i\eta )[\exp \beta(\gamma E_k'+\gamma vk_3'-\mu)-1]}}.
\nonumber
\end{eqnarray}
The step function factor in the first term imposes $q_3<0$ so that the 
assumed asymptotic behavior holds at the pinch.  Rewrite these terms to 
exhibit the poles in the $k_3$ plane.
\begin{eqnarray}
\lefteqn{S_A(k_\perp,k_3,k_3')\sim}\\
\nonumber\\
&&{{|t_P(\gamma k_3'-\gamma vE_k')|^2\theta (\gamma v
\sqrt{k_\perp^2+m^2}-k_3')}\over
{2\pi i(k_3-k_3'-i\eta )[\exp \beta(\gamma E_k'
-\gamma vk_3'-\mu)-1]}}
\nonumber\\
\nonumber\\
\nonumber\\
&&-{{|t_P(\gamma k_3'+\gamma vE_k')|^2|r_V(k_3')|^2}\over
{2\pi i(k_3-k_3'+i\eta )[\exp \beta(\gamma E_k'+\gamma vk_3'-\mu)-1]}}.
\nonumber
\end{eqnarray}
The first of these poles is in the first quadrant, and contributes when the 
contour is deformed to $C$, but the second pole is in the fourth quadrant 
and is irrelevant.

     When all contributions are added, and the unitarity relations invoked, 
the relevant singularity of $S_A$ is
\begin{equation}
S_A(k_\perp,k_3,k_3')\sim{{{\rm sgn}(k_3')}\over{2\pi i [k_3-k_3'-i\eta
{\rm sgn}(k_3')][\exp\beta(\gamma E_k'-\gamma vk_3'-\mu)-1]}}.
\end{equation}
There is no relevant pole at $k_3=-k_3'$.  The term $S_A$ makes 
the time-independent contribution
\vfill
\eject
\begin{displaymath}
J_A\sim\int{{d^3k}\over{2(2\pi )^3E}}\left [{\cal Z}_V^*(k_3,z)
\left (-i{{\stackrel{\leftrightarrow}d}\over{dz}}
\right ){\cal Z}_V(k_3,z)\right ]{1\over{[\exp\beta (\gamma E-\gamma 
vk_3-\mu)-1]}}
\end{displaymath}
\begin{equation}
\sim\int{{d^3kk_3|t_V(k_3)|^2}\over{(2\pi )^3E}}{1\over{[\exp\beta 
(\gamma E-\gamma vk_3-\mu)-1]}}.
\end{equation}

     In the term proportional to $S_B$, deform the $k_3$-integration from 
the real axis to contour $C'$, which lies in the second and fourth quadrants 
of the complex $k_3$-plane.  There is then a time-independent contribution 
from the following pole in $S_B$:
\begin{eqnarray}
\lefteqn{S_B(k_\perp,k_3,k_3')}\\
&&\sim-{{{\rm sgn}(k_3')}\over{2\pi i [k_3-k_3'+i\eta
{\rm sgn}(k_3')]}}\left \{{1\over{[\exp\beta(\gamma E_k'-\gamma vk_3'
+\mu)-1]}}+1\right \}.\nonumber
\end{eqnarray}
The addition to the baryon current is
\begin{equation}
J_B\sim-\int{{d^3kk_3|t_V(k_3)|^2}\over{(2\pi )^3E}}
\left \{{1\over{[\exp\beta (\gamma E-\gamma vk_3+\mu)-1]}}+1\right \}.
\end{equation}
The extra term in the brace is present because we did not normal order the 
current operator.  No matter: The resulting integrand is odd in $k_3$ and 
vanishes.  Altogether, the baryon current in the limit $t\rightarrow
\infty$ is
\begin{displaymath}
J=\int{{d^3kk_3|t_V(k_3)|^2}\over{(2\pi )^3E}}
\left \{ {1\over{\exp\beta (\gamma E-\gamma vk_3-\mu)-1}}\right.
\end{displaymath}
\begin{equation}
\label{eq:freecurrent}
-\left.{1\over{\exp\beta (\gamma E -\gamma vk_3+\mu)-1}}\right \}.
\end{equation}

     This result is expected.  The Bose distributions are those of a 
moving free gas; they give the densities of baryons and antibaryons, 
which differ for $\mu\ne 0$.  The contributions to the net current 
density are opposite for baryons and antibaryons.  They are given by 
particle density multiplied by velocity, $k_3/E$, and the quantum 
mechanical probability of transmission past the VEV wall.

\section{Plasmas with Self-Interactions}
\setcounter{equation}{0}

     The result of the last Section shows that the approach based on 
statistical mechanics works for a plasma without self-interactions, for 
any velocity of the plasma relative to the VEV wall.  The methods used 
there relied on the absence of self-interactions.  
On the other hand, the physical question of the influence of quasiparticle 
lifetime on the baryon current can be addressed only when we take 
self-interactions into account in a systematic way.  This can be done 
by reformulating the problem as a standard problem in many-body theory.  
In particular, we want to reduce the problem to the computation of a 
temperature Green's function.

     In this Section, we so recast the problem, but only to first order 
in $v$.  Note that the current of Eq.~\ref{eq:freecurrent} is odd in $v$: 
$J(-v)=-J(v)$.  We expect this reflection property to hold in a 
self-interacting theory also, when the VEV wall profile in the vicinity 
of $z=L$ is the mirror image of the profile in the vicinity of $z=0$.  
Then the correction to our $O(v)$ result is $O(v^3)$, and the $O(v)$ 
result is applies over a useful range of $v$.

     Write the statistical matrix of the self-interacting plasma in the 
form
\begin{equation}
\rho=e^{-\beta K_P}=e^{-\beta\gamma K_V}U(\beta );\qquad U(\tau)=e^
{\tau\gamma K_V}e^{-\tau K_P}.
\end{equation}
Introduce the boost operator, and its ``imaginary time Heisenberg'' picture:
\begin{equation}
B=\gamma K_V-K_P;\qquad B(\tau )=e^{\tau\gamma K_V}Be^{-\tau\gamma K_V}.
\end{equation}
Then
\begin{eqnarray}
{{dU(\tau )}\over{d\tau}}&=&B(\tau )U(\tau );\\
U(\tau )&=&\sum_{n=0}^\infty{1\over{n!}}\int_0^\tau d\tau_1\cdots d\tau_n
T_\tau\left [B(\tau_1 )\cdots B(\tau_n )\right ],\nonumber
\end{eqnarray}
where in the product, larger $\tau$ operators stand to the left of smaller 
$\tau$ operators.  $\gamma K_V$ and $K_P$ have the same self-interactions, 
so
\begin{equation}
\label{eq:boost}
B=-\gamma v\int d^3x\left [{{\partial\hat\phi^\dagger}\over{\partial z}}
\hat\pi^\dagger +\hat\pi{{\partial\hat\phi}\over{\partial z}}\right ].
\end{equation}
The power series for $U$ is therefore equivalent to an expansion in powers of 
$v$.  Through first order in $v$ (but all orders in self-interactions)
\begin{equation}
JTr[\rho ]=Tr[e^{-\beta K_V}\hat J_3({\bf x},t)]+\int_0^\beta d\tau 
Tr[e^{-\beta K_V} B(\tau )\hat J_3({\bf x},t)]+\dots .
\end{equation}
The first term on the right is the baryon current of a plasma at rest in 
frame $V$; it is zero.  The second term is the first order baryon current 
$J(1,{\bf x},t)$:
\begin{equation}
\label{eq:cons}
J(1,{\bf x},t)Tr[\rho ]=\int_0^\beta d\tau Tr[e^{-\beta K_V} B(\tau )\hat 
J_3({\bf x},t)].
\end{equation}

     The Lehmann representation for $J(1,{\bf x},t)$ is~\cite{Lehmann}
\begin{eqnarray}
J(1,{\bf x},t)Tr[\rho ]=\sum_{mn}\int dE_mdE_ne^{i(E_m-E_n)t}\left [
{{e^{-\beta (E_n-N_m\mu)}-e^{-\beta (E_m-N_m\mu)}}\over{E_m-E_n}}\right ]\\
\times\int d\sigma_md\sigma_n\langle m|\hat J_3({\bf x})|n\rangle\langle n|
B|m\rangle .\nonumber
\end{eqnarray}
The sum is over eigenstates of $H_V$ and $N$.  It is partially discrete and 
partially an integral over $dEd\sigma$; the state energy $dE$ is exhibited.  
We expect the inner, partially integrated, expression to include a 
contribution proportional to an energy delta function:
\begin{equation}
\int d\sigma_md\sigma_n\langle m|\hat J_3({\bf x})|n\rangle\langle n|B
|m\rangle =\delta (E_m-E_n)\langle m|{\cal J}|n\rangle +\dots
\end{equation}
The delta function contribution arises because operator $B$ is an integrated 
density.  It is present for a free plasma, as we will 
show below, and it is there in each order in the perturbative 
expansion in the plasma self-interaction.  Furthermore, current conservation 
requires $\langle m|{\cal J}|n\rangle$ to be independent of $z$, because 
$E_m=E_n$.  Between eigenstates of $H_V$, current conservation states
\begin{equation}
0=\langle m |{{\partial\hat J_0}\over{\partial t}}
+{\bf\nabla\cdot\hat J}|n\rangle =\langle m |i(E_m-E_n)\hat J_0+{\bf\nabla
\cdot\hat J}|n\rangle .
\end{equation}
When $E_m=E_n$, this becomes
\begin{equation}
0={\bf\nabla\cdot}\langle m|{\bf\hat J}|n\rangle ={\partial\over{\partial z}}
\langle m|\hat J_3|n\rangle .
\end{equation}
Only the energy-conserving contribution survives in the limit $t\rightarrow
\infty$, leaving the finite $z$-independent current
\begin{equation}
\label{eq:firstord}
J(1)Tr[\rho ]=J(1,{\bf x},t=\infty)Tr[\rho ]=\left.\beta\sum_{m,n}\int dE_m
\langle m|{\cal J}|n\rangle\right |_{E_n=E_m}.
\end{equation}

     We can relate $J(1)$ to the imaginary time temperature Green's function
\begin{eqnarray}
{\cal G}({\bf x},\tau_1;\tau_2)Tr[\rho ]&=&-Tr\left \{e^{-\beta K_V}T_{\tau}
\left [\hat J_3({\bf x},\tau_1 )B(\tau_2)\right ]\right \},\nonumber\\
\hat J_3({\bf x},\tau)&=&e^{\tau K_V}\hat J_3({\bf x})e^{-\tau K_V}.
\end{eqnarray}
This Green's function has familiar properties: It is a function of $\tau_1-
\tau_2$, and is periodic in this variable with period $\beta$.  The Fourier 
transform $\tilde{\cal G}({\bf x},\omega_p)$, with $\omega_p=2\pi p/\beta$, is
\begin{equation}
\tilde{\cal G}({\bf x},\omega_p)=\int_0^\beta d\tau e^{i\omega_p\tau}{\cal G}
({\bf x},\tau ;0).
\end{equation}
This Fourier transform is the amplitude that has a simple expansion in 
Feynman diagrams in many-body perturbation theory.  Its Lehmann representation 
is
\begin{eqnarray}
\label{eq:imagtime}
\tilde{\cal G}({\bf x},\omega_p)Tr[\rho ]=-\sum_{mn}\int dE_mdE_n\left [
{{e^{-\beta (E_n-N_m\mu)}-e^{-\beta (E_m-N_m\mu)}}\over{E_m-E_n+i\omega_p}}
\right ]\\
\times\int d\sigma_md\sigma_n\langle m|\hat J_3({\bf x})|n\rangle\langle n|
B|m\rangle .\nonumber
\end{eqnarray}
We see immediately that $-\tilde{\cal G}({\bf x},\omega_p=0)$ is identical 
with $J(1,{\bf x},t=0)$.  However, Eq.~\ref{eq:imagtime} has contributions 
from states with $E_m\ne E_n$ that make it differ from $J(1)$, 
Eq.~\ref{eq:firstord}.  Fortunately, there is a simple way to eliminate 
the contributions from unwanted states and recover $J(1)$.  We see from 
Eq.~\ref{eq:cons} that states with $E_m=E_n$ make contributions to $\tilde
{\cal G}({\bf x},\omega_p=0)$ that are independent of $z$, while the 
unwanted states make contributions that depend on $z$.  There is no 
difficulty in extracting the $z$-independent contribution; it pops out when 
the integration in Eq.~\ref{eq:boost} is carried out.  Thus the relation 
between $J(1)$ and the temperature Green's function is
\begin{equation}
J(1)=-{\cal G}({\bf x},\omega_p=0)|_{z-{\rm ind}}.
\end{equation}

     Note that no analytic continuation in $\omega$ is required in this 
problem, in contrast to the situation when one studies real-time transport 
phenomena at finite temperature.

     We are primarily interested in $\tilde{\cal G}({\bf x},\omega_p=0)$ when 
the plasma has self-interactions.  However, it is straightforward to compute 
it for a free plasma.  It is useful to do so, because the emergence of 
the energy conserving delta function is illustrated, and the correctness of 
the result can be confirmed.  We find
\begin{displaymath}
-{\cal G}({\bf x},\omega_p=0)=-v\int{{d^2k_\perp dk_3dk'_3}\over
{4(2\pi )^4EE'}}\left [{\cal Z}_V^*(k_3,z)\left (-i{{\stackrel
{\leftrightarrow}d}\over{dz}}\right ){\cal Z}_V(k'_3,z )\right ]
\end{displaymath}
\begin{eqnarray}
\lefteqn{\times\left \{{{e^{\beta (E-\mu)}-e^{\beta (E'-\mu)}}\over
{(E-E')[e^{\beta (E-\mu )}-1][e^{\beta (E'-\mu )}-1]}}\right.}\nonumber\\
& &-\left.{{e^{\beta (E+\mu)}-e^{\beta (E'+\mu)}}
\over{(E-E')[e^{\beta (E+\mu )}-1][e^{\beta (E'+\mu )}-1]}}\right \}
\end{eqnarray}
\begin{displaymath}
\times\int_{-\infty}^\infty dz'\left [-iE{{d{\cal Z}_V^*(k'_3,z')}\over{dz'}}
{\cal Z}_V(k_3,z')+iE'{\cal Z}_V^*(k'_3,z'){{d{\cal Z}_V(k_3,z')}\over{dz'}}
\right ]
\end{displaymath}
\begin{displaymath}
-v\int{{d^2k_\perp dk_3dk'_3[e^{\beta (E-\mu)}e^{\beta (E'+\mu )}-1]}
\over{4(2\pi )^4EE'(E+E')[e^{\beta (E-\mu )}-1][e^{\beta (E'+\mu )}-1]}}
\end{displaymath}
\begin{eqnarray}
\times\left \{\left [{\cal Z}_V^*(k_3,z)\left (-i{{\stackrel
{\leftrightarrow}d}\over{dz}}\right ){\cal Z}^*_V(k'_3,z )\right ]
\int_{-\infty}^\infty dz'\left [-iE{{d{\cal Z}_V(k'_3,z')}\over{dz'}}
{\cal Z}_V(k_3,z')\right.\right.\nonumber\\
\left.-iE'{\cal Z}_V(k'_3,z'){{d{\cal Z}_V(k_3,z')}\over{dz'}}
\right ]\nonumber\\
+\left [{\cal Z}_V(k_3,z)\left (i{{\stackrel
{\leftrightarrow}d}\over{dz}}\right ){\cal Z}_V(k'_3,z )\right ]
\int_{-\infty}^\infty dz'\left [iE{{d{\cal Z}^*_V(k'_3,z')}\over{dz'}}
{\cal Z}^*_V(k_3,z')\right.\nonumber\\
\left.\left.+iE'{\cal Z}^*_V(k'_3,z'){{d{\cal Z}^*_V(k_3,z')}\over{dz'}}
\right ]\right \}.\nonumber 
\end{eqnarray}
First note that this Green's function does not depend on ${\bf x_\perp}$.  
The reason is that the transverse integrations in Eq.~\ref{eq:boost} set 
${\bf k_\perp}={\bf k'_\perp}$.  However, $k_3\ne k'_3$ because of the 
VEV wall, and the Green's function certainly depends on $z$.  
The integrations over $z'$ come from Eq.~\ref{eq:boost}.  They do not 
impose momentum or energy conservations generally, but the asymptotic tails 
of these integrals do produce energy conserving delta functions in addition 
to other contributions.  Such delta functions lead to discrete 
position-independent contributions to ${\cal G}({\bf x},
\omega_p=0)$.  The delta function terms occur in just one of the integrals:
\begin{eqnarray}
\int_{-\infty}^\infty dz'\left [-iE{{d{\cal Z}_V^*(k'_3,z')}\over{dz'}}
{\cal Z}_V(k_3,z')+iE'{\cal Z}_V^*(k'_3,z'){{d{\cal Z}_V(k_3,z')}\over{dz'}}
\right ]\\
=-4\pi k_3E\left [|t_V(k_3)|^2\delta (k_3-k_3')+r^*_V(-k_3)t_V(k_3)\delta 
(k_3+k_3')\right ]+\dots.\nonumber
\end{eqnarray}
Retaining just the delta functions, the $z$-independent contribution to 
\hfill
\break
${\cal G}({\bf x},\omega_p=0)$ implies
\begin{eqnarray}
\lefteqn{J(1)=-{\cal G}({\bf x},\omega_p=0)|_{z-{\rm ind}}}\\
& & =v\beta\int{{d^3kk^2_3|t_V(k_3)|^2}\over{(2\pi)^3E}}\left \{
{{e^{\beta (E-\mu)}}\over{[e^{\beta (E-\mu )}-1]^2}}-{{e^{\beta (E+\mu)}}
\over{[e^{\beta (E+\mu )}-1]^2}}\right \}.\nonumber
\end{eqnarray}
This agrees with the $O(v)$ baryon current obtained by expanding Eq.~\ref
{eq:freecurrent}.  

     Physical questions about the baryon current are thus 
converted into questions about a temperature Green's 
function.  We have experience and techniques for the study of such objects.

\bigskip

\noindent{\large{\bf Acknowledgements}}

This work was supported in part by the National Science Foundation under
grant number NSF-PHY-94-23002.

\bibliographystyle{plain} 
\vfill
\eject

\end{document}